\documentstyle[12pt]{article}

\newcounter{multieqs}

\newcommand{\bq}{\begin{equation}}
\newcommand{\fq}{\end{equation}}
\newcommand{\bqr}{\begin{eqnarray}}
\newcommand{\fqr}{\end{eqnarray}}
\newcommand{\non}{\nonumber \\}

\newcommand{\xpv}[1]{\langle #1  \rangle}

\newcommand{\rf}[1]{(\ref{#1})}

\def\npb#1#2#3{Nucl. Phys. {\bf{B#1}} (#2) #3}
\def\plb#1#2#3{Phys. Lett. {\bf{#1B}} (#2) #3}

\def\prd#1#2#3{Phys. Rev. {D \bf{#1}} (#2) #3}

\def\ap#1#2#3{Annals of Physics {\bf{#1}} (#2) #3}

\def\jhep#1#2#3{J. High Energy Phys. {\bf #1} (#2) #3}
\def\atmp#1#2#3{Adv. Theor. Mat. Phys. {\bf #1} (#2) #3}
\def\alp{\alpha}       
\def\del{\delta}   \def\eps{\epsilon} 
\def\zet{\zeta}        
    \def\kap{\kappa}   \def\lam{\lambda}
 \def\sig{\sigma}   
 \def\vphi{\varphi} \def\ome{\omega}

\def\Gam{\Gamma}   \def\Del{\Delta}   
\def\Lam{\Lambda}

\def\cM{{\cal M}} \def\cN{{\cal N}} \def\cO{{\cal O}}

\def\pa{\partial}

\def\dag{^{\dagger}}
\def\pr{^{\prime}}

\def\rar{\rightarrow}

\def\one{1\!\!1\,\,}

\newcommand{\tr}{\mbox{Tr}}

\def\ove#1{\frac{1}{#1}}

\def\bk{{\vec{k}}}
\begin{document}

\thispagestyle{empty}

\marginparwidth = .5in

\marginparsep = 1.2in

\begin{flushright}
\begin{tabular}{l}
ANL-HEP-PR-98-116 \\ 
ITP-SB-98-59 \\
\\
hep-th/9810051
\end{tabular}
\end{flushright}

\vspace{18mm}
\begin{center}

{\bf The large $N_c$ limit of four-point functions in $N=4$
super-Yang-Mills theory from anti-de Sitter Supergravity}

\vspace{18mm}

{Gordon Chalmers${}^{\dagger}$ and Koenraad Schalm${}^*$
 }\footnote{E-mail addresses: chalmers@sgi2.hep.anl.gov and
konrad@insti.physics.sunysb.edu}
\\[10mm]
${}^{\dagger}${\em Argonne National Laboratory \\
High Energy Physics Division \\
9700 South Cass Avenue \\
Argonne, IL  60439 } \\[5mm]
${}^*${\em Institute for Theoretical Physics  \\
State University of New York       \\
Stony Brook, NY 11794-3840 }

\vspace{20mm}

{\bf Abstract}

\end{center}

We compute the imaginary part of scalar four-point functions in the
AdS/CFT correspondence relevant to $N=4$ super Yang-Mills theory.
Unitarity of the AdS supergravity demands that the
imaginary parts of the correlation functions factorize into
products of lower-point functions. We include the exchange diagrams
for scalars as well as gravitons and find explicit expressions for the
imaginary parts of these correlators. In momentum space these
expressions contain only rational functions and logarithms of the
kinematic invariants, in such a manner that the correlator is not a
free-field result.  The simplicity of these results, however, indicate
the possibility of additional symmetry structures in $N=4$ super
Yang-Mills theory in the large $N_c$ limit at strong effective
coupling.  The complete expressions may be computed from the integral
results derived here.

\vfill

\setcounter{page}{0}
\newpage
\setcounter{footnote}{0}

\baselineskip=16pt

\section{Introduction}

In the past few months much work has focused on the
correspondence between superconformal field theory
and string (M-)theory in an
anti-de Sitter background \cite{mal}.  In particular, the low energy
supergravity regime of the compactified $\mbox{AdS}_{p+1}$
string theory should correspond to the large $N_c$ limit of a
$p$-dimensional conformal field theory at strong 't Hooft coupling
$\lam=g^2_{YM}N_c$.  A holographic relation 
between conformal field theory correlators and those in the 
gauged supergravity is given in 
\cite{gub,wit}:  correlations in the CFT are
generated via,
\bq
\prod_{j=1}^k \left.\left( {\delta\over \delta\phi_{0,j}(\vec{z}_j)}
\right) e^{iS_{\rm sugra}[\phi(\phi_0)]}\right|_{\phi_{0,j}=0} =
\langle
\prod_{j=1}^k {\cal O}^j(\vec{z}_j)
\rangle_{CFT}
\ .
\label{fundamental}
\fq
$S_{sugra}$ is the bulk action of the compactified string theory
considered as a functional of the boundary values of the fields,
$\phi_{0,j}$, and ${\cal O}$ are composite (gauge
invariant) operators  of the conformal Yang-Mills theory. These
operators are dual to the boundary values of the supergravity
fields in the sense that the latter act as sources
for the former, as can be seen from
\rf{fundamental}.

A specific example is the correspondence between $d=4$
$N=4$ super-Yang-Mills theory in the large $N_c$ limit at large $\lam$
and type IIB supergravity on $\mbox{AdS}_5 \times S_5$ with radii
$R_{AdS}^2=R_{S_5}^2 = \alpha' \sqrt{4\pi g_{\rm st} N_c}$, where
$g_{\rm YM}^2= 4\pi g_{\rm st}$.  It is clear that testing
the conjecture is not an easy task as the accessible regimes for
computations, $\lam \ll 1$ for super-Yang-Mills theory and
$\lam \gg 1$ for supergravity, do not overlap.  Nevertheless two- and
three-point correlators of certain chiral primary operators and their
descendants have been calculated and the supergravity results were found to
agree with the free-field results on the super-Yang-Mills side
\cite{freed1,tseyt1,us,min,howewest}.  For two-point functions this was
expected due to the known existence of a  non-renormalization
theorem.  For certain three-point correlators a similar theorem was
suspected to exist and in \cite{min} this was conjectured
for all three-point functions of chiral
operators from explicit computation. Such a
non-renormalization theorem should follow from the
stringent covariance constraints of the $N=4$
superconformal algebra.  Recent 
evidence has been given in~\cite{freed4}.

Superconformal constraints are not expected to protect
the four-point or higher order correlation functions from
renormalization.  In particular, it is not clear that free-field
Yang-Mills result will be reproduced by the supergravity
calculation at the four-point level.  Evidence for this claim 
is the fact that certain
four-point functions have non-trivial $\lambda$-dependence.
They are affected by $\alp\pr$ string corrections to
the classical supergravity action, which correspond
to $\lam^{-1/2}$ terms in the strong ('t Hooft)
coupling expansion of the CFT
\cite{banks,sav,brogut}.  Thus, an  explicit comparison at $n \geq 4$
order is a non-trivial test of  the dynamics of $N=4$ super
Yang-Mills theory.
In view of the above, any indications of further
non-renormalization  theorems for four-point
functions of particular operators would be very
interesting on the field theory side.

In this article we will calculate the imaginary part of four-point
functions in AdS supergravity using the
unitarity constraints on Green's functions.  Unitary
quantum field theories obey the identity,
\bq
2\mbox{Im}\, T = T\dag T
\label{t-ident}
\fq
where $T$ is the transition matrix $iT=S-\one$.
These unitarity relations hold also for general Green's functions and
not just the case where the external lines are on-shell.
Classical supergravity, as well as finite-$N_c$ super
Yang-Mills theory, is certainly unitary and we will
use this to relate the imaginary parts of four-point functions to the
product of three-point boundary-boundary-bulk functions, whose explicit
expressions can be computed.  Explicit computations of four-point functions
are technically challenging; however, following the unitary cutting
approach we avoid the technical complications associated with
integrating over the bulk-bulk propagator.  Though we limit ourselves to
the imaginary parts in particular channels our approach enables us to
make predictions regarding the infinite summation of planar graphs.
Specifically, we may compare with the generically complicated
logarithmic dependence of the perturbative series in field theory.
(The complete correlators may also be found from our computations by
performing an additional one-parameter integral and adding diagrams
containing four-point vertices.)

This work is organized as follows. In section 2 we formulate the
holographic
Feynman rules in Lorentzian signature AdS used to
generate the boundary correlators. We find the
bulk-bulk and bulk-boundary kernels for scalars and
gravitons used in the following sections.  The bulk
vertices are also found.  We then discuss unitarity
principles in section 3 and explain the cutting
rules. In section 4 we derive the imaginary parts of
the correlator of four axions in IIB
supergravity on $\mbox{AdS}_5 \times S_5$ and show how the calculation
reduces to the square of three-point functions; in section 5 we
discuss how other scalar correlators may be computed. In section 6 we
interpret the
result within the AdS/CFT correpondence for the $N=4$
super-Yang-Mills theory.

Previous to this work contributions to four-point functions
involving four-point vertices were given in \cite{muck}.  Scalar
exchange contributions were discussed in \cite{tseyt2,freed2},
and recently the expression for the contribution due to the
exchange of a vector boson in a covariant
gauge was obtained  \cite{covgauge}.

\section{Kernels in AdS${}_{d+1}$}
\setcounter{equation}{0}

We need to formulate the supergravity theory in a Lorentzian signature
version of anti-de Sitter space in order to derive the imaginary parts
within a given channel. Most of the work relating to
the AdS/CFT correspondence has been performed using the Euclidean version
described in $\cite{wit}$.\footnote{There is a family of AdS quantum
vacua with Euclidean signature preserving the isometry structure
\cite{allen}.} Our prescription will be to naively Wick rotate from
Euclidean AdS to a Lorentzian form.  Some care has to be taken in 
Lorentzian signature as one has to confront the
subtlety that there are additional homogenous  solutions  to
the field equations \cite{vijay1}; the bulk supergravity modes are no
longer uniquely determined from their boundary  values. We will
comment on this below.

In Euclidean space the bulk-bulk propagator for a given supergravity field
may be written as
\bq
\Del(x,y) = -\sum_n \frac{\vphi^{\ast}_n(y)\vphi_n(x)}{\lam_n^2}
\label{prop}
\fq
where the $\vphi_n(x)$ span a complete set of
eigenfunctions of the kinetic operator that obey the correct
boundary conditions. The summation extends over all quantum numbers,
denoted by
$\{n\}$, necessary to describe the solution and the $\lam^2_n$ are
the eigenvalues associated with $\vphi_n(x)$.  In  Wick rotating to
Lorentzian signature one needs to provide an $i\eps$ prescription,
$\lam_n^2 \rar\lam_n^2 \pm i\eps$, in the denominator of~\rf{prop}.
The presence (as well as the sign) of the $i\eps$ term is determined
by the physical requirement that at large timelike separations
the positive frequency modes dominate.  

We shall consider the halfspace Poincar\'{e} coordinate system for
$\mbox{AdS}_{d+1}$ with metric
\bq
ds^2 = \frac{A^2}{x_0^2}\left(dx_0^2 + dx^idx^j \eta_{ij}\right),~~
i=1,...,d
\label{met}
\fq
where $x_0 \geq 0$. The metric $\eta_{ij}$ is Minkowski with
mostly plus signature and $\eta_{dd}= -1$, and the analytical continuation
of the metric to its Euclidean form is obvious.  The boundary of AdS in
these coordinates is the Minkowski space ${\mbox{\bf R}}^{3,1}$ at
$x_0=0$ and the single point $x_0=\infty$. This metric has a timelike
Killing vector $\pa/\pa x^d$ whose conserved charge we may interpret as
the energy and according to which an $i\eps$ precription may be
given \cite{freed3,isham}.

This section is devoted to constructing the eigenfunctions
$\vphi_n(x)$ for the scalar and graviton fields in the AdS
background. In this article we will consider a
particular set of boundary conditions (Dirichlet) for the bulk AdS
modes. This uniquely selects a complete set of modes $\vphi_{n}(x)$
for the Euclidean kernel. After the Wick rotation some
of these may become zero modes but all have to be retained
in the kernel to preserve completeness.  Besides such zero modes there
are further ones which alter the form of
the correlators and their boundary behaviour when added to the
propagator \cite{vijay1}. These
are necessary to describe the AdS theory with another choice of
boundary conditions for the bulk fields. In view of unitarity and the
dependence in the imaginary parts this is interesting in that there
may be choices where particular cuts vanish.  Here we will
take the conventional route by not adding these solutions, which is
analogous to a vacuum choice \cite{vijay1, vijay2}.

We will concern ourselves with scalar four-point functions in the
dilaton-axion sector of IIB supergravity on
$\mbox{AdS}_5\times S_5$. It was shown in \cite{tseyt2} that at the level of four-point fucntions this sector of the compactified supergravity theory does not receive contributions from any other fields. The CFT operators which correspond to these
fluctuations are $\tr F^2$ and  $ \tr F\tilde{F}$ for the dilaton and
axion respectively. The relevant part of the IIB action
is given by\footnote{Our
conventions for the Riemann and the Ricci tensor are that
$R_{\mu\nu\rho}^{~~~~\tau} = \pa_{\mu}\Gam_{\nu\rho}^{~~\tau} +
\Gam_{\mu\sig}^{~~\tau}\Gam^{~~\sig}_{\nu\rho} - (\mu
\leftrightarrow \nu)$ and
$R_{\mu\rho} = R_{\mu\nu\rho}^{~~~~\nu}$.}
\bqr
S= {1\over 2\kap_{d+1}^2} \int d^{d+1}x \sqrt{g} \Bigl[
-({{R-\Lam}}) -{1\over 2} g^{\mu\nu} \partial_\mu \phi \partial_\nu
\phi -{1\over 2} e^{2\phi} g^{\mu\nu}\partial_\mu C\partial_\nu C
\Bigr]
\label{stringframe}
\ ,
\fqr
where $d=4$ and $1/2\kap_5^2=\Omega_5 N_c^2 /(2\pi)^5= N_c^2/15\pi^3$
in terms of the $N=4$ super-Yang-Mills variables. The proportionality
of the supergravity boundary
values to the CFT operators may be fixed by a two-point function
calculation.  As in \rf{stringframe} the
correspondence between supergravity and super-Yang-Mills theory is
usually formulated in the Einstein frame.  The exact
relation between the five-dimensional supergravity fields and the
conformal operators is not precisely clear because of subtleties
in the consistent truncation from the Kaluza-Klein modes.
The cosmological constant
$\Lambda$ is related to the scale $A$ of the Poincar\'{e} metric as
\bq
\Lam = \frac{d(d-1)}{A^2} \ .
\fq
Below we will set the scale $A$ to unity, in which case the scalar
curvature associated with the background AdS metric equals $R =
d(d+1)$.

\subsection{Scalars}

We first recall the derivation of the propagator for a
scalar field on anti-de Sitter space. Along the way we find the
correct $i\eps$ prescription needed in Lorentzian signature.

The covariant kinetic operator for massive scalars in a curved
background is
\bqr
\hat{K} \Phi(x) &=& \left(\ove{\sqrt{g}}\left(\pa_{\mu} \sqrt{g}
g^{\mu\nu} \pa_{\nu}\right) -m^2 \right) \Phi(x) \non
&=&  \left(x_0^{d+1}\pa_0\left(\ove{x_0^{d-1}}\pa_0\right) +
x^2_0\pa^i\pa_i - m^2\right)\Phi(x) \ .
\label{scalarop}
\fqr
The differential operator in \rf{scalarop} has the
eigenfunctions \cite{tseyt2}
\bqr
\vphi_{\lam}(x) &=& x_0^{d/2} e^{i\vec{k} \cdot \vec{x}}J_{\nu}(\lam x_0),
 ~~~\vec{k} \cdot \vec{x} \equiv \sum_{i=1}^d k_ix^i
\non
\hat{K} \vphi_{\lam}(x) &=& -(\lam^2+\vec{k}^2)x_0^2 \vphi_{\lam}(x) \
,
\fqr
where $\nu =\sqrt{m^2+d^2/4} >0$. The eigenfunctions
$\vphi_{\lam}(x)$ are labelled by the four-vector $\vec{k}$, the
conserved momentum along the boundary, and a continous eigenvalue
$\lam$. The other Bessel functions that are possible solutions
of~\rf{scalarop} do not obey the Dirichlet boundary condition at
$x_0=0$ or are not well-behaved in the interior of AdS.

The Bessel functions $J_{\nu}(\lam x_0)$  and the Fourier modes
are complete in the sense that
\bq
\int_0^\infty d\lam\, \lam J_{\nu}(\lam x_0) J_{\nu}(\lam y_0) =
\frac{\del(x_0-y_0)}{\sqrt{x_0y_0}} \ ,
\fq
and
\bq
 \int \frac{d^dk}{(2\pi)^d}
e^{i\vec{k} \cdot (\vec{x}-\vec{y})} = \del^d(\vec{x}-\vec{y}) \ .
\fq
The Euclidean bulk-bulk propagator for scalars is thus given by
\bq
G_{\Phi}(x,y) \equiv \xpv{\Phi(x)\Phi(y)} = \int_0^{\infty} d\lam \,
 \lam \int \frac{d^dk}{(2\pi)^d}
 \frac{\vphi^{\ast}_{\lam}(x)\vphi_{\lam}(y)}{(\lam^2+\bk^2)} \ ,
\label{scBBprop}
\fq
and obeys
\bq
\hat{K} \xpv{\Phi(x)\Phi(y)} = - x_0^{d+1}\del^{d+1}(x-y) =
-\frac{\del^{d+1}(x-y)}{\sqrt{g}} \ .
\fq
To obtain the Lorentzian signature version of the propagator we
analytically continue the $d$th coordinate. Requiring that the
positive frequency modes $k^{Lor}_d = -\ome(\bk) <0$ dominate at large
times we find that we should effectively replace $(k_d^E)^2$ with
$-(k_d^{Lor})^2 - i\eps$ in~\rf{scBBprop}.

In order to compute the conformal field theory correlation
functions from the supergravity diagrams we also need the bulk-boundary
kernel for the scalars:
\bqr
\Del(\vec{x},y) &=& \sqrt{h}n^{\mu}\pa_{\mu}G_{\Phi}(x,y)|_{x \eps
\pa\cM} \ ,
\non
\phi_{bulk}(y) &= &\int d^dx {\Del}(\vec{x},y) \phi_{bound}(\vec{x}) \
,
\fqr
where $h$ is the determinant of the induced metric on the boundary.
These are attached as the
external legs of the supergravity Green's function, according to the
prescription of ~\cite{gub,wit}. Care has to be taken in defining the
bulk-boundary propagator, since formally the Poincar\'{e} metric blows
up at the boundary \cite{wit,freed1}.
One way to regularise is to impose the
Dirichlet conditions at
$x_0=\zet$ and let
$\zet
\rar 0$ at the end of the calculation. The accordingly modified
Green's function is \cite{freed1,muck}
\bq
G^{\zet}_{\Phi}(x,y) = G_{\Phi}(x,y) - (x_0y_0)^{d/2}\int
\frac{d^dk}{(2\pi)^d}e^{i\vec{k}
\cdot(\vec{x}-\vec{y})}
K_{\nu}(kx_0)K_{\nu}(ky_0) \frac{I_{\nu}(k\zet)}{K_{\nu}(k\zet)} \ ,
\fq
where $k=\sqrt{\vec{k}\cdot \vec{k}}$.  Taking the normal
derivative then yields
\bqr
\Del^{\zet}(\vec{x},y) &=& \left(\frac{y_0}{\zet}\right)^{d/2} \int
\frac{d^dk}{(2\pi)^d} e^{i\vec{k} \cdot(\vec{x}-\vec{y})}
\frac{K_{\nu}(ky_0)}{K_{\nu}(k\zet)} \ .
\fqr
We have not examined the $i\epsilon$ procedure of the bulk-boundary 
kernels in order to determine cuts in single-particle channels here, 
although one may certainly do so.

For massless scalars the index of the Bessel function $J_{\nu}(\lam
x_0)$ equals $\nu=d/2$ and the limit of $\zet \rar 0 $
can be taken without ambiguity. Hence for this special case the
bulk-boundary propagator is regularisation
independent and equals
\bq
{\Del}(\vec{x},y)|_{\nu=d/2} =
\frac{2}{\Gam(d/2)} \int
\frac{d^dk}{(2\pi)^d} ~\left(\frac{k y_0}{2}\right)^{d/2}K_{d/2}(k y_0) 
{e^{i\vec{k}\cdot
 (\vec{x}-\vec{y})}} \ .
\fq
We will not apply our methods in this work to the evaluation of
boundary correlators for massive scalars; however, the techniques
presented below are certainly available for this case as well.

\subsection{Gravitons in $h_{\mu 0}=0$ Gauge}

For simplicity we have chosen to work in the $h_{\mu 0}=0$ gauge. Of
course this choice breaks the background isometries of the AdS space
and is therefore not preferred above a covariant gauge such as
$g^{\mu\nu}D_{\mu}h_{\nu\rho}=0$. However, the kinetic operators in
the latter gauges are not readily inverted due to the
presence of the nontrivial background curvature (see e.g.
\cite{gravprop}).   The external supergravity fields satisfy 
field equations near the boundary of the AdS space.  This indicates  
that final results for correlator expressions should in principle 
be independent of the gauge-choice, although this has not yet 
been examined in detail.

Expanding the action for the graviton,
$\hat{g}_{\mu\nu}\rightarrow { g}_{\mu\nu} + h_{\mu\nu}$, in the
presence of a source term $T_{\mu\nu}$,
\bq
S= {1\over 2\kappa^2_{d+1}}
  \int d^{d+1}x \left(\sqrt{\hat{g}(x)} ~ \left[-( R- \Lambda) \right]+
\sqrt{g(x)}h_{\mu\nu} T^{\mu\nu}\right)  \ .
\label{actionnnn}
\fq
we find the effective field equations for $h_{\mu\nu}$,
\bqr
&&\ove{2}\left\{g^{\mu\tau}g^{\nu\sig}g^{\rho\alp}(D_{\rho}D_{\alp}h_{\tau\sig}
-D_{\rho}D_{\tau}h_{\sig\alp}
-D_{\rho}D_{\tau}h_{\alp\sig})\right.
\non&&\hspace{.2in}\left.
+g^{\mu\nu}g^{\rho\alp}g^{\sig\tau}D_{\rho}D_{\sig}h_{\alp\tau}
+g^{\mu\tau}g^{\nu\sig}D_{\tau}D_{\sig}h
-g^{\mu\nu}g^{\rho\alp}D_{\rho}D_{\alp}
h\right\}
\\&&\nonumber\hspace{.9in}
-\ove{2}g^{\mu\tau}g^{\nu\sig}h_{\tau\sig}(R-\Lam)
+\ove{4}g^{\mu\nu}R^{\rho\sig}h_{\rho\sig}
+\ove{4}R^{\mu\nu}h=
-T^{\mu\nu} +\cO(h_{\mu\nu}^2)\ .
\fqr
The indices are raised with the background AdS metric
${g}^{AdS}_{\mu\nu}$ and the derivatives $D_{\alp}$ are covariant
with respect to this metric.

Setting $h_{\mu 0}=0$, we find the field equation
\bqr
\pa_0^2(x_0^2h_{mn}-\eta_{mn}(x_0^2h))
-\frac{(d-1)}{x_0}\pa_0(x_0^2h_{mn}-\eta_{mn}(x_0^2h)) \non
+x_0^2\left(\Box\tilde{h}_{mn}-2\pa_m\pa_i\tilde{h}_{in}
+\eta_{mn}\eta^{ij}\pa_i\pa_k\tilde{h}_{jk}\right)
=-\frac{2}{x_0^4}T_{mn} \ ,
\label{knk}
\fqr
where $\tilde{h}_{mn}=h_{mn}-\ove{2}\eta_{mn}h$ and
$h=\eta^{mn}h_{mn}$.  The box is defined as $\Box=\partial_i^2$ with
flat $d$-dimensional indices.  The field equations for $h_{0\mu}$
generate  constraints,
\bqr
\pa_0(\pa_m (x_0^2h)-\pa_i (x_0^2h_{im}))
=-\frac{2}{x_0^4}T_{m}^0 \ ,
\label{conone}
\fqr
and,
\bqr
 \eta^{ij}\pa_i\pa_k
(x_0^2h_{jk})-\Box(x_0^2h)+\frac{(d-1)}{x_0}\pa_0(x_0^2h) =-
\frac{2}{x_0^4}T^{00} \ .
\label{contwo}
\fqr
In equations \rf{knk} through \rf{contwo} boundary indices have
been raised and lowered with $\eta_{ij}$.
Following \cite{tseyt2} we decompose $h_{mn}$ into,
\bqr
&&h_{mn}=h^{\perp}_{mn}+\pa_mV^{\perp}_n+\pa_nV_m^{\perp} +\pa_m\pa_nS
+
\ove{(d-1)}(\del_{mn}-\frac{\pa_m\pa_n}{\Box})h\pr \ , \non&&
\hspace{1in}
\eta^{mn}h^{\perp}_{mn}=\pa^mh^{\perp}_{mn}=\pa^mV^{\perp}_m = 0 \ .
\fqr
Projecting the field equation~\rf{knk} onto the boundary transverse
traceless part we obtain
\bq
\left(x_0^2\pa_0^2+x_0^2\Box -(d-5)x_0\pa_0-2(d-2)\right)
h^{\perp}_{mn} = -\frac{2}{x_0^4}t_{mn}~,~~~t_{mn}=P_{mnij}T_{ij} \ ,
\fq
where $P_{mnij}$ is the transverse traceless projector on the
boundary flat Minkowski space.  The propagator for the fluctuation
$x_0^2h^{\perp}_{mn}$ is then the same as that of a massless
scalar.

The remaining unphysical components are constrained by \rf{conone}
and \rf{contwo},
\bqr
\pa_0(x_0^2V^{\perp}_i)&=&
\frac{2}{x_0^4}\ove{\Box}\left(\del_{ij}-\frac{\pa_i\pa_j}{\Box}\right)T^0_j
\ ,\\
\pa_0(x_0^2h\pr) &=& -\frac{2}{x_0^4}\ove{\Box}\pa_jT^0_j \ , \\
\pa_0(x_0^2S) &=&
\frac{x_0}{(d-1)}\left(-\frac{2}{x_0^4}\ove{\Box}T^{00}+x_0^2h\pr\right)
+\frac{2}{x_0^4}\ove{\Box^2}\pa_jT^0_j\ .
\fqr
We then use the fact that $T^{\mu\nu}$ is conserved,
\bqr
D_{\mu}T^{\mu\nu}=0  &\rar&\left\{
\begin{array}{c}
 \pa_iT^{ij} = -x_0^{d+3}\pa_0x_0^{-d-3}T^{0j} \\[0.1in]
\pa_iT^{i0} = -x_0^{d+2}\pa_0x_0^{-d-2}T^{00}-\ove{x_0}T^{jj}
\end{array}
\right.
\fqr
and use these relations in the  action.  Integrating twice
by parts we find the source action (up to boundary terms which do not
contribute to bulk propagation) \cite{tseyt2},
\bqr
S
&=& {1\over 2\kappa^2_{d+1}} \int d^{d+1}y \sqrt{g(y)} ~ d^{d+1}z
\sqrt{g(z)} ~ t^{ij}(y) 2\eta_{ik}\eta_{j\ell}G_{h^{\perp}}(y,z)
t^{k\ell}(z)
\non &&
-{1\over 2\kappa^2_{d+1}}\int d^{d+1}x \sqrt{g(x)}~
\Bigl[  T_{0i} {4\over
x_0^6}\ove{\Box}
T_{0i}
\non&&  +T^{00}\frac{4}{(d-1)x_0^5}\ove{\Box}\pa_jT^{0j}
- \pa_jT^{0j}\frac{2(d-2)}{(d-1)x_0^6}\ove{\Box^2} \pa_iT^{0i}
\Bigr]
\ .
\label{gravlight}
\fqr
Explicitly the Green's function for the physical modes $h^{\perp}_{ij}$
is :
\bqr
G_{h^{\perp}}(x,y) &=& \ove{(x_0y_0)^2}G_{\Phi,m^2=0}(x,y)
\non
&=& (x_0 y_0)^{d/2-2} \int d\lambda \lambda
 \int {d^d k\over (2\pi)^d} ~ {J_{d/2} (\lambda x_0) J_{d/2}(\lambda y_0)
\over \lambda^2+\vec{k}^2-i\epsilon} e^{i\vec{k}\cdot (\vec{x}-\vec{y})} \ .
\fqr

One may also find the propagator for vectors in the $A_0=0$ gauge in a
similar way. In that case one obtains the source action \cite{tseyt2}.
\bq
S=\ove{2}\int d^{d+1}y \sqrt{g(y)} ~ \int d^{d+1}z \sqrt{g(z)}
~~  {\cal J}^{\perp,i}\eta_{ij} G_{A}(y,z) {\cal
J}^{\perp,j}
\fq
\bq
-{1\over 2} \int d^{d+1}x \sqrt{g(x)}~ {\cal J}_0 {1\over \Box}
{\cal J}_0 \ ,
\label{aogauge}
\fq
The correlator of physical polarizations is  proportional to
the propagator of a massive field with mass $m^2=1-d$
\bqr
G_A(x,y) &=& \ove{(x_0y_0)} G_{\Phi,m^2=1-d}(x,y)
\non &=&
(x_0 y_0)^{d/2-1} \int d\lambda \lambda
 \int {d^d k\over (2\pi)^d} ~ {J_{\rho} (\lambda x_0) J_{\rho}(\lambda y_0)
\over \lambda^2+\vec{k}^2-i\eps} e^{i\vec{k}\cdot (\vec{x}-\vec{y})} \
,
\fqr
where $\rho^2 = (1-d)+d^2/4$ (or $\rho^2=1$ when $d=4$).

A brief comment is in order about the apparent
additional poles in the $A_0=0$ and $h_{0\mu}=0$ gauge fixed 
form of the bulk action.  (The effective action
will in general be gauge dependent but gauge invariant when the 
external sources do not themselves satisfy field equations, 
for example in the background field method.)  Related to these 
gauge fixed actions is the light-cone form of 
Yang-Mills theory, in the spinor notated form,
\bqr
{\cal L} = {\rm Tr} ~~
 {\bar A} \Box A + \partial_{+\dot +} {\bar A} \Bigl[
\left(\partial^\alpha{}_{\dot +} {1\over \partial_{+\dot +}} A \right)
\left(\partial_{\alpha \dot +} {1\over \partial_{+\dot +}} A \right)
 \Bigr] + {\rm c.c.} + {\cal O}(A^2,{\bar A}^2) \ ,
\fqr 
which has a similar structure as the gauge fixed actions above,
and is readily available to compute gauge invariant S-matrices as well as 
effective actions (including its $N=4$ extension).  Time 
derivatives are $\partial_{+\dot +}$.  In gauge invariant expressions 
the potential poles are spurious.

Within the imaginary parts of the correlator involving
non-covariantly gauge fixed intermediate states, the
poles in the gauge fixed action in the latter terms of \rf{gravlight} do
contribute.  They lead to contributions in the holographic
Feynman diagrams with a $\Box-i\epsilon$ in the denominator and
contribute to a cut only when $(\vec{k}_i+\vec{k}_j)^2=0$. Presumably,
these contact interactions exist to cancel spuriously introduced
infra-red divergent terms.

\section{Unitarity and Cutting}
\setcounter{equation}{0}

We shall adopt the unitarity cutting procedure to the supergravity
theory in order to compute the four-point (and higher) correlators.
In flat space the imaginary part in a particular channel
is read off by complex conjugation through the identity,
\bq
\ove{x+i\eps}-\ove{x-i\eps} = 2\pi i \del(x)
\label{iden}
\fq
Holding, for example, in the momentum space expression the
external momenta $k_i^2>0$ and analytically continuing
$s_{12}$ to negative values will pick out the unitarity cut in this
channel.  If we denote the generic four-point correlator,
$C(\vec{k}_j)$ and its complex conjugate $C^\star(\vec{k}_j)$; the
imaginary part
$C-C^\star$, after imposing $s_{12}<0$ and $k_j^2>0$ receives a
contribution only from cutting the intermediate propagator in the
$s_{12}$ channel via the relation in \rf{iden}.

To see the unitarity relations in its simplest form consider a scalar
field theory in Minkowski space with a cubic interaction. The
propagator for this field is given by the usual
\bq
\xpv{\vphi(x)\vphi(y)} = \int\frac{d^4k}{(2\pi)^4}
\frac{e^{ik(x-y)}}{k^2+m^2-i\eps}
\fq
The tree level $ 2 \rar2 $ scattering diagram is thus given
by two vertices sandwiching a propagator. The
 imaginary part of this graph in the two-particle channel is found
through the use of~\rf{iden};  we immediately see that the graph splits
in the product of two three point graphs where the intermediate state
is on shell.  Note, however, that the external lines are {\em not}
required to be on shell, but only have to satisfy momentum
conservation.

The supergravity result should yield the
leading strong effective coupling term in the planar diagram sector
of the theory.  In gauge theory at
one-loop the reduced tensor integrations in supersymmetric theories
enable one to find the complete correlator (and effective actions)
from explicit information of only the imaginary parts in all channels
\cite{bdkone,bdktwo}.\footnote{This is even the case for
non-supersymmetric theories, provided one keeps the full form of the
dimensionally regulated integral  functions.}  The
extraction of the imaginary part requires  performing the phase space
integral,
\bqr
{\rm Im}\vert_{s} ~A_{n;1} (k_1,\ldots,k_n) &=&
\sum_{\lambda_1,\lambda_2} ~ \int d\phi_2~ A_{n;1}^{\rm tree}(k_1,\ldots,
p_1^{\lambda_1},\ldots,p_2^{\lambda_2}\,\ldots,k_n)
\non &&
 \times  A_{n;1}^{\rm tree}(k_1,\ldots, p_1^{-\lambda_1},\ldots,
 p_2^{-\lambda_2},\ldots,k_n) \ ,
\label{cut}
\fqr
where $d\phi_2$ is the two-particle phase space measure,
$\delta(p_1^2) \delta(p_2^2) \Theta(p_1^0)\Theta(p_2^0)$, and
we have suppressed  color structures.
Extracting the full amplitude follows from the logarithmic
orthogonality of the one-loop integral reduction formulae
\cite{bdkone,bdktwo}.

At two or more loops the cuts in
a particle channel require
higher than two-particle cuts; for example, one may take a
cut of a double box through a two-particle or three-particle
channel.  Such generalizations of the program to finding
complete correlators (and S-matrix elements) from the imaginary
parts, as at one-loop, have only been partially developed 
\cite{multiloop}.

In the previous section we introduced the necessary $i\eps$ terms in the
Lorentzian formulation of the AdS supergravity theory. The delta-function
identity~\rf{iden} then relates
the imaginary part of the four-point function in the two-particle channels
to two three-point boundary-boundary-bulk functions, the functional form
of which may be explicitly calculated. Different from conventional
field theory is that the prescription of
\cite{gub,wit} instructs us to work with Green's functions with
bulk-boundary kernels attached to the external legs.  

\section{Factorization}
\setcounter{equation}{0}

We shall illustrate our technique with a calculation involving
four scalar fields, axions, of IIB supergravity on $\mbox{AdS}_5 \times S_5$.
The general four-point scalar function involves graphs built with a
bulk four-point vertex , which do not contribute to the imaginary
parts in the $s_{ij}$ channels, and graphs with two three-point
vertices joined by an intermediate scalar or graviton line. We
shall here consider in detail the
$\langle C(\vec{x}_1) C(\vec{x}_2) C(\vec{x}_3) C(\vec{x}_4)
\rangle$ correlator dual to the the $N=4$ SYM four-point function
$\xpv{\prod_{i=1}^4 \tr F\tilde{F}(\vec{x}_i)}$ as
it is the simplest to analyze.  In this case there are no four-point
vertices and only six holographic diagrams of the second type
described above, each with an internal line associated with a scalar or
graviton in the $s$,
$t$, or
$u$ channel.

\subsection{Scalar Exchange}

We first examine the contribution to this correlator coming from
the dilaton exhange.  The $\phi(x)C(x)C(x)$ vertex is,
\bq
{\cal L}_{\phi CC} = -\ove{2\kap^2} \sqrt{g} g^{\mu\nu}
\phi
\partial_\mu  C \partial_\nu C \ ,
\fq
and hence the expression for the correlator arising 
from an internal
dilaton ($\phi$) line is,
\bqr
A_{CCCC}^{\phi,s}(\vec{x}_i) &=& \frac{4\cdot 2}{2!}\ove{2\kap^2}\int
d^{d+1}z_1
\sqrt{g(z_1)} ~\int d^{d+1}z_2
\sqrt{g(z_2)}
\non &\times &\left[
g^{\mu\nu}(z_1) \partial_\mu \Del(\vec{x}_1,z_1) \partial_\nu
\Del(\vec{x}_2,z_1) \right] G_{\Phi,m^2=0}(z_1,z_2)
\non &\qquad &
\qquad\left[
g^{\alpha\beta}(z_2) \partial_\alpha \Del(\vec{x}_3,z_2)
\partial_\beta
 \Del(\vec{x}_4,z_2) \right]\ ,
\fqr
together with the $t$- and $u$-channel diagrams.
After Fourier transforming with respect to the directions parallel
to the boundary
\bq
A_{CCCC}^{\phi,s}({\vec k}_i) = \Bigl( \prod_{j=1}^4 \int
{d^{d}x_i}
 e^{i\vec{k}_j \cdot \vec{x}_j} \Bigr) ~
A_{CCCC}^{\phi,s}(x_1,x_2,x_3,x_4) \ .
\fq
we find the momentum space expression for the $s$-channel
contribution to the correlator\footnote{Note that since the 
imaginary part is nonzero the above diagram does not reduce 
solely to that of an
effective four-point vertex. Previous
articles \cite{tseyt2,freed2} used partial integration of the bulk derivatives
to relate scalar exchange diagrams to an effective four-point vertex.
The integration, however, produces boundary terms that contribute to cuts. 
}
\bqr
A_{CCCC}^{\phi,s}(\vec{k}_i) &=&
\frac{2}{\kap^2}\delta^{(d)}(k_1+k_2+k_3+k_4)
\int dy_0 y_0^{-d+1} ~
\int dz_0 z_0^{-d+1}
\non &\times&
I({\vec k}_1,{\vec k}_2,y_0)~I({\vec k}_3,{\vec k}_4,z_0)
\non &\times&
\int d\lambda {\lambda\over
\lambda^2+(\vec{k}_1+\vec{k}_2)^2-i\epsilon}
 (y_0z_0)^{d/2} J_{d/2}(\lambda y_0) J_{d/2}(\lambda z_0)
  \ ,
\label{scalar}
\fqr
where,
\bqr
I({\vec k}_1,{\vec k}_2,y_0) &=& -\vec{k}_1\cdot \vec{k}_2~
\Del(k_1,y_0) \Del (k_2,y_0)
\non &&
\qquad
 + \partial_{y_0} \Del(k_1,y_0)
 ~\partial_{y_0}  \Del (k_2,y_0)\ .
\fqr

The imaginary part of the expression in \rf{scalar}, when
holding $\vec{k}_i^2>0$ and $s=(\vec{k}_1+\vec{k}_2)^2<0$,
receives a contribution equal to the effective replacement
of the bulk-bulk propagator with
\bq
-\pi \int d\lambda ~\lambda ~\delta(\lambda^2+\vec{k}^2)
~(y_0 z_0)^{d/2}~ J_\nu(\lambda y_0) J_\nu(\lambda z_0)  \ .
\fq
The $\lambda$ integral may be performed and we shuffle the factor
$y_0^{d/2} J_\nu(\lambda_0 y_0)$ to one side of the cut and
the $z_0$-term to the other side (the additional factor of $\lambda$ is
cancelled by the delta function).  We also have
$\lambda=\pm i\sqrt{\vec{k}^2}$,  with $\vec{k}=\vec{k}_1+\vec{k}_2$.

In general then, for any massive intermediate field (or gauge field for
that matter), the cut of the four-point holographic Feynman diagram
reduces to the product of two boundary-boundary-bulk three-point
functions. In the  case above, in \rf{scalar}, we have the example of
an internal  scalar and hence
\bq
{\rm Im}~ A_{CCCC}^{\phi,s}(\vec{k}_j)  = -\frac{\pi}{\kap^2}
\delta^{(d)}(k_1+k_2+k_3+k_4)  ~ M(\vec{k}_1,\vec{k}_2,\lam)
M(\vec{k}_3,\vec{k}_4,\lam)|_{\lam^2=-\vec{k}^2} \ ,
\fq
where $\vec{k}=\vec{k}_1+\vec{k}_2$ and
\bqr
M(\vec{k}_1,\vec{k}_2,\lam) &=& \frac{2^{2-d}}{\Gam^2(d/2)}\int dy_0
y_0^{d/2+1} ~
  J_{d/2}(\lam y_0)
\non &\times&
\Bigl[ -\vec{k}_1\cdot \vec{k}_2~ (y_0k_1)^{d/2} (y_0k_2)^{d/2}
 K_{d/2}(y_0k_1) K_{d/2}(y_0k_2)
\non &&
\quad
+ \partial_{y_0} \bigl\{ (y_0k_1)^{d/2} K_{d/2}(y_0k_1) \bigr\}
 ~\partial_{y_0} \bigl\{ (y_0k_2)^{d/2} K_{d/2}(y_0k_2) \bigr\}
\Bigr] \ .
\label{halfcut}
\fqr
The second integral in \rf{halfcut} may be simplified by noting
that
\bq
\partial_{y_0} \bigl\{ (y_0k_i)^{d/2} K_{d/2}(y_0k_i) \bigr\}
= k_i{\partial\over\partial k_i} \Bigl\{ {1\over y_0}
(y_0k_i)^{d/2} K_{d/2}(y_0k_i) \bigr\} \ ,
\label{blurb}\fq
and then extracting the momentum derivatives outside the
integration.

To proceed further, we need the expression for a triple
convolution of Bessel functions (related integrals are given 
in \cite{bw}).
In order to evaluate integrals of the general form,
\bq
B^{t}_{\nu_1,\nu_2,\nu_3} = \int_0^{\infty}dy ~y^{t+1}
K_{\nu_1}(k_1y)K_{\nu_2}(k_2y) J_{\nu_3}(\lam y) \ ,
\label{triple}
\fq
we use the following integral expression for the modified
Bessel function $K_\nu(x)$:
\bq
K_{\nu}(\lam x)  = \ove{2} \left({2x\over \lam}\right)^\nu
\int_0^{\infty}
 d\tau \, \tau^{\nu-1} e^{-x^2 \tau - \frac{\lam^2}{4\tau}} \ .
\label{Kintegral}
\fq
We would also like to point out that in three-point
correlation function calculations of bilinear operators, the $\tau$ parameters of the
bulk-boundary kernels when expressed through \rf{Kintegral} become
Schwinger parameters of triangle integrals \cite{us} (offering a partial
explanation that these correlations are derived in field theory
at one-loop).

Using \rf{Kintegral} the triple convolution in \rf{triple} may be
expressed as
$$
B^{t}_{\nu_1,\nu_2,\nu_3} = {1\over 4}
  \left( {4\over k_1^2}\right)^{\nu/2}
  \left( {4\over k_2^2}\right)^{\nu/2}
\int \prod_{j=1}^2 d\tau_j ~\tau_j^{\nu_j-1} e^{-k_j^2/4\tau_j}
$$
\bq
\times
\int_0^{\infty}dy ~y^{t+1+\nu_1+\nu_2} e^{-(\tau_1+\tau_2)y^2}
 J_{\nu_3}(\lam y) \ .
\fq
The $y$-integration may now be done, and it gives the
formal expression \cite{bateman},
$$
\int_0^{\infty}dy ~y^{t+1+\nu_1+\nu_2} e^{-(\tau_1+\tau_2)y^2}
 J_{\nu_3}(\lam y) = {1\over 2} {\Gamma({1\over 2}[\nu_1+\nu_2 +
\nu_3 +t] +1)\over \Gamma(\nu_3+1)}
$$
\bq
\qquad
\times
\left[{\lam^2\over 4(\tau_1+\tau_2)} \right]^{\nu_3/2}
 {}_1F_1\Bigl[1+{1\over 2}(\nu_1+\nu_2+\nu_3+t);\nu_3+1;
 -{\lam^2\over 4(\tau_1+\tau_2)}\Bigr]  \ ,
\fq
where ${}_1 F_1$ is a confluent hypergeometric function.
Its defining series representation,
\bq
{}_1F_1[a;b;x] = \sum_{n=0}^\infty {\Gamma(a+n)\over\Gamma(a)}
{\Gamma(b)\over\Gamma(b+n)} {x^n\over n!} \ ,
\fq
with the condition that $a\geq b$, permits us to write it in
the form,
\bqr
{}_1 F_1(a;b;x) &=& \frac{\Gam(b)}{\Gam(a)}P_{a-b}(x)e^x
\label{newcon}
\fqr
The $P_{a-b}(x)$ is a polynomial of degree $a-b$, related to the
associated Laguerre polynomials $L^k_n(x)$, with leading unit
coefficient,
\bqr
P_{a-b}(x) &=& x^{a-b} + \gamma_0 x^{a-b-1} + \ldots +
\gamma_{a-b}  \non &=& \Gam(a-b+1) e^xL_{a-b}^{b-1}(-x)\ .
\fqr
For $a-b ~\eps~ \mbox{\bf Z}^+$, the operator
$P_{a-b}(\frac{\pa}{\pa\mu})$ has the simple representation
\bq
P_{a-b}(\frac{\pa}{\pa\mu})f(\mu)=\left(\frac{\pa}{\pa\mu}\right)^{a-b}
\left[ \mu^{a-1}f(\mu) \right] \ .
\fq
In the
case $a-b$ is non-integral we may formally regard the operator
as an infinite series expansion. However, all the cases analyzed
in detail here will pertain to integral values of $a-b$.

The
re-expression of the confluent function
${}_1 F_1$ in \rf{newcon} allows us to complete the $\tau_i$
integrals.
The integral $B^{t}_{\nu_1,\nu_2,\nu_3}$ after extracting the
polynomial derivative $P_{a-b}$ is,
$$
B^{t}_{\nu_1,\nu_2,\nu_3}={1\over 8} \left( {4\over k_1^2}\right)^{\nu_1/2}
 \left( {4\over k_2^2}\right)^{\nu_2/2}
 \left( {\lam^2\over 4}\right)^{\nu_3/2}
P_{{1\over 2}(\nu_1+\nu_2+t-\nu_3)}({\pa\over \pa\mu})
$$
\bq
\times
\int \prod_{j=1}^2 d\tau_j ~\tau_j^{\nu_j-1} e^{-k_j^2/4\tau_j}
~ (\tau_1+\tau_2)^{-{1\over 2}(\nu_1+\nu_2+\nu_3+t+2)}
 e^{-\mu \lam^2/4(\tau_1+\tau_2)} \ .
\fq
Now switching
integration variables first to,
\bq
\tau_i \rightarrow {1\over \tau_i} \qquad d\tau_i \rightarrow
-{1\over \tau_i^2} d\tau_i \ ,
\fq
and then to
\bq
\tau_1=\alpha \tau \qquad \tau_2=(1-\alpha)\tau \qquad
d\tau_1 d\tau_2 = \tau d\tau d\alpha \ ,
\fq
we may also complete the integral over $\tau$,
\bqr
B^t_{\nu_1,\nu_2,\nu_3} &=&
\frac{2^{t-1}\lam^{\nu_3}}{k_1^{\nu_1}k_2^{\nu_2}}\Gam(a-\nu_1-\nu_2)
\left(\frac{\pa}{\pa \mu}\right)^{a-b} \mu^{a-1}
\non &&
\quad \times \int_0^1 d\alp~
\frac{\alp^{a-\nu_2-1}(1-\alp)^{a-\nu_1-1}}
{[k_1^2(1-\alp)+k_2^2\alp +
\mu\lam^2\alp(1-\alp)]^{a-\nu_1-\nu_2}} \vert_{\mu=1} \ ,
\fqr
with
\bq
2a = \nu_1+\nu_2+\nu_3+t+2 ~~,~~ b=\nu_3+1 \ .
\fq
In the case of massless
scalar exchange, we have $\nu_1=\nu_2=\nu_3=t=d/2$ and the remaining
$\alp$ integration is finite; we  set  $d=4$ as well.
Substituting these values the last integral may be performed, and
yields
\bqr
B^2_{2,2,2} &=& -\frac{2}{k_1^2k_2^2}\left(\frac{\pa}{\pa\mu}\right)^2
~{\mu^3}\left\{ \ove{3} -\frac{3}{2} (q_++q_-)+(q_+^2+q_+q_-+q_-^2) \right.
\non
&&\left.\left.
+\frac{q_+^2(1-q_+)^2}{(q_+-q_-)}\ln\left[\frac{1-q_+}{q_+}\right]
-\frac{q_-^2(1-q_-)^2}{(q_+-q_-)}\ln\left[\frac{1-q_-}{q_-}\right]
\right\}\right|_{\mu=1}
\label{genint}
\fqr
where $q_\pm$ are the roots of the quadratic,
\bq
\alpha^2 - {k_1^2-k_2^2+\mu \lam^2\over \mu \lam^2} \alpha
- {k_2^2\over \mu \lam^2} =0 \ ,
\fq
or,
\bqr
q_{\pm} &=& \frac{(k_1^2-k_2^2+\mu \lam^2)}{2\mu \lam^2}
\pm \frac{\sqrt{(k_1^2-k_2^2+\mu \lam^2)^2+4\mu
k_2^2 \lam^2}}{2\mu \lam^2} \ .
\label{quad}
\fqr
It is noteworthy that the functional form of \rf{genint}
contains only rational functions in the external momenta and logarithms.

Similarly, the second integral in \rf{halfcut} has $\nu_1=\nu_2=\nu_3
=d/2$ and $t=d/2-2$.  This integral is also finite and may be evaluated
 to:
$$
k_1{\partial\over\partial k_1}
k_2{\partial\over\partial k_2} ~B^{d/2-2}_{d/2,d/2,d/2}
= 32 k_1^2 k_2^2 \lam^2 P_1({\pa\over \pa\mu}) \bigl[ {\pa\over
\pa(\mu \lam^2)}
\bigr]^2
$$
\bq
\times \Bigl\{ \ln(-\mu k_3^2) + \sum_{\pm}~ \ln(1-q_\pm)
+ q_\pm \ln(-{q_\pm\over 1-q_\pm}) \Bigr\} \vert_{\mu=1} \ ,
\fq
with $d=4$.  Alternatively, one may use the recursion relation
\bq
\ove{x^{\nu}}\pa_x(x^{\nu}K_{\nu}(\lam x)) = -\lam K_{\nu-1}(\lam x) \ ,
\label{blab}
\fq
instead of \rf{blurb}. The second contribution to
$M(\vec{k}_1,\vec{k}_2,\lam)$ is then proportional to,
\bqr
B^2_{1,1,2}
&=&
\frac{2}{\lam^2\sqrt{k_1^2 k_2^2} }
\left(\frac{\pa}{\pa \mu}\right)~
\frac{\mu}{(q_+-q_-)^2}\left\{ \ove{q_+(q_+ -1)}+\ove{q_-(1-q_-)} \right.
\non
&&\left.\left.
-\frac{2}{(q_+ -q_-)}\ln\left[\frac{1-q_+}{q_+}\right]
+\frac{2}{(q_+ -q_-)}\ln\left[\frac{1-q_-}{q_-}\right]
\right\}\right|_{\mu=1} \ .
\fqr

The cut in the $s$-channel of the $A_{CCCC}^{\phi}$-correlation
function arising from the intermediate dilaton  exchange is then,
\bq
{\rm Im}~ A_{CCCC}^{\phi,s}(\vec{k}_j)  = -\frac{\pi}{\kap^2}
\delta^{4}(k_1+k_2+k_3+k_4)  ~
M(\vec{k}_1,\vec{k}_2,\lam)
M(\vec{k}_3,\vec{k}_4,\lam)|_{\lam^2=-\vec{k}^2}
\ ,
\fq
with
\bqr
M(\vec{k}_1,\vec{k}_2,\lam)&=& \left[ -k_1\cdot k_2 B^2_{2,2,2}
+
 k_1\frac{\pa}{\pa k_1}k_2\frac{\pa}{\pa k_2} B^{0}_{2,2,2}\right]
\non
&=& \left[ -k_1\cdot k_2 B^2_{2,2,2}
+
 \sqrt{k_1^2 k_2^2} B^{2}_{1,1,2}\right] \ .
\fqr
Similar manipulations may
be used to find the contribution from massive states as well.

We end this section by noting how the entire exchange diagram
with the intermediate scalar may be found.  Effectively in the
above we have integrated completely the fifth coordinates $y_0$
and $z_0$ laying at the ends of the bulk-bulk propagator.
We may alternatively use these integrals to reconstruct the full
diagram through one more additional integral (i.e. the one over $\lambda$),
\bq
 A_{CCCC}^{\phi,s}(\vec{k}_j) =
\frac{2}{\kap^2}\del^d(\vec{k}_1+\vec{k}_2+\vec{k}_3+\vec{k}_4)
\int_0^\infty
d\lambda ~
 M(\vec{k}_1,\vec{k}_2,\lambda) {\lambda
 \over \lambda^2+ (\vec{k}_1+\vec{k}_2)^2}M(\vec{k}_3,\vec{k}_4,\lambda) \
.
\label{fullcorr}
\fq
The $\lambda$ integration
is over a product of logarithms and rational
functions, and it is reasonable to suspect that this last integration
 may be done generally and explicitly.

\subsection{Graviton Exchange in $h_{\mu 0}=0$ Gauge}

In this gauge the contribution of the intermediate graviton
is also easily computed.  The bulk-bulk Green's function for the
physical polarizations is that of a massless scalar, and the
calculation is nearly the same as in the previous section.

The coupling of the axion fields to the graviton is derived by
expanding the quadratic term
\bqr
{\cal L}_{hCC} &=& - {1\over 4\kap^2} \sqrt{g} g^{\mu\nu} \partial_\mu
C
\partial_\nu C +
 {1\over 2\kap^2} \sqrt{g}~ h_{\mu\nu} T_C^{\mu\nu}
\label{stress}
\fqr
about the anti-de Sitter background $g_{\mu\nu}$.  This defines the
energy-momentum tensor induced from the scalar matter,
\bq
2T_C^{\mu\nu}(z) = \partial^\mu C(z) \partial^\nu
C(z) - {1\over 2} g^{\mu\nu} \partial^\alpha C(z) \partial_\alpha C(z)
 \ .
\fq
Using the induced action in \rf{gravlight} we
may derive the $s$-cut in the graviton exchange diagram.

The projection onto the transverse traceless components $t_{ij}$ gives
\bqr
t^{ij}(z) &=&  \frac{x_0^4}{2}P^{ij,mn} \left[ \partial_m C \partial_n
C - {1\over 2} g_{mn} \partial^\alpha C \partial_\alpha C \right]
\non &
=&  \frac{x_0^4}{2}P^{ij,mn} \partial_m C \partial_n C
 \ ,
\fqr
with which we may compute the exchange diagram
coming from the physical components of the graviton. Recall that $P^{ij,mn}$
is the transverse traceless projector on the $d$-dimensional boundary.

The first term in \rf{gravlight} then gives the contribution,
\bqr
A_{CCCC}^{h^{\perp},s}(x_i) =\frac{4\cdot 2}{2!}\ove{(2\kap^2)} \int
d^{d+1}y\sqrt{g(y)}~
\int d^{d+1}z\sqrt{g(z)} ~~ \left[\pa_{m} \Del(x_1,y) \pa_{n}
\Del(x_2,y)\right] \non
\times ~P^{mn,rs}  (y_0 z_0)^4  G_{h^{\perp}}(y,z)
 \left[ \pa_{r} \Del(x_3,z) \pa_{s} \Del(x_4,z)\right] \ ,
\fqr
Hence we find the contribution in the $s$-channel,
\bqr
{\rm Im}~A_{CCCC}^{h^{\perp},s} = -\frac{\pi}{\kap^2}
\delta^{(d)} (k_1+k_2+k_3+k_4) ~
M^{ab}(\vec{k}_1,\vec{k}_2,\lam) \left.
M_{ab}(\vec{k}_3,\vec{k}_4,\lam)
 \right|_{\lam^2=-\vec{k}^2}\ ,
\fqr
Immediate application of the preceding techniques gives us ,
\bq
M^{ab}(\vec{k}_1,\vec{k}_2,\lam) =
-P^{ab,mn}\vec{k}_{1,m}\vec{k}_{2,n}B^{d/2}_{d/2,d/2,d/2}
\fq
In $d=4$ the integral is finite and has been done before; we only quote
the result,
\bqr
B^2_{2,2,2} &=& -\frac{2}{k_1^2k_2^2}\left(\frac{\pa}{\pa\mu}\right)^2
~{\mu^3}\left\{ \ove{3} -\frac{3}{2} (q_++q_-)+(q_+^2+q_+q_-+q_-^2) \right.
\non
&&\left.\left.
+\frac{q_+^2(1-q_+)^2}{(q_+-q_-)}\ln\left[\frac{1-q_+}{q_+}\right]
-\frac{q_-^2(1-q_-)^2}{(q_+-q_-)}\ln\left[\frac{1-q_-}{q_-}\right]
\right\}\right|_{\mu=1} \ .
\fqr
Again the final form is such that there are only logarithms and
rational functions contributing.

The remaining diagrams are generated by interactions resembling
four-point vertices; however, from the $i\epsilon$ contained
in the transverse flat box, $\Box-i\epsilon$, there are unitarity
cuts only when the on-shell condition $(\vec{k}_i+\vec{k}_j)^2=0$
is satisfied.  Although we shall not consider these terms in any
more detail, these integrals may be evaluated
and are necessary in the construction of the full correlator.

\section{Other Correlator examples}
\setcounter{equation}{0}

Although we have examined a few of the diagrams involved
in the general correlator, the methods (and momentum space
integration techniques) generalize to other correlator 
expressions.

Let us consider for example the correlator of two dilatons and two
axions,  i.e. $\langle \phi(\vec{k}_1) C(\vec{k}_2)
\phi(\vec{k}_3) C(\vec{k}_4) \rangle$.
The main difference between this example and those in preceeding
sections involves the derivative structure
acting on the intermediate axion line joined between two
three-point vertices:
\bqr
A_{\phi C\phi C}^{C,s}(\vec{x}_j)
 &=& \frac{2}{2!}\ove{(2\kap^2)}\int \sqrt{g(z_1)} d^{d+1}z_1 ~ \int
\sqrt{g(z_2)} d^{d+1}z_2
\non&&
g^{\alpha\beta}(z_1) \left[ \Del(\vec{x}_1,z_1) \partial_\beta^{(1)}
\Del(\vec{x}_2,z_1)
\right]g^{\mu\nu}(z_2) \left[ \Del(\vec{x}_3,z_2) \partial_\mu^{(2)}
\Del(\vec{x}_4,z_2)
\right] \non
&&\times \partial_\alpha^{(1)}
\partial_\nu^{(2)}G_{\Phi,m^2=0}(z_1,z_2)
\label{threephipiece} \ .
\fqr
The  derivative with respect to $y_0$ on the (bulk-)Bessel function
$J_{\nu}(\lam z_i)$ can be evaluated using either the recursion
relation in~\rf{blurb} or~\rf{blab}. Alternatively one may partially
integrate this derivative away onto the bulk-boundary propagator
carrying no derivative. The boundary term vanishes due to the Dirichlet
condition on the bulk to bulk Green's function and the remaining
terms produced by the partial integration conspire to\footnote{This 
is a special case; in general boundary terms do not vanish 
\cite{holn}.}
\bq
\ove{\sqrt{g(z_i)}}\left(\pa_{\mu}^{(i)} \sqrt{g(z_i)}
g^{\mu\nu}(z_i) \pa_{\nu}^{(i)}\right) \Del(\vec{x},z_i) =0 \ .
\fq
The integral is then the same as in~\rf{scalar}.

We see that again the imaginary part of these expressions
factorizes into two identical functions.
\bq
{\rm Im} ~A_{\phi C\phi C}^{C,s}(\vec{k}_j)
=
-\frac{\pi}{4\kap^2}\delta^{(d)}(\vec{k}_1+\vec{k}_2+\vec{k}_3+\vec{k}_4)
~
 \left.M(\vec{k}_1,\vec{k}_2,\lam)
M(\vec{k}_3,\vec{k}_4,\lam)\right|_{\lam^2=-\vec{k}^2}\ .
\fq
This factorization occurs for
{\it all} diagrams because the correlation functions are generated
through supergravity tree diagrams.  In principle, one may explicitly
compute this way the contributions from the exchange of massive scalars
and other tensor fields, once the propagators are known.
The general structure of the imaginary part of any correlator in
momentum space, following from the generic structure of intermediate
propagators in \rf{prop}, is thus
\bq
{\rm Im} ~C(\vec{x}_j)
= \sum_{I}~ \delta^{(d)}(\vec{k}_1+\vec{k}_2+\vec{k}_3+\vec{k}_4) ~
\left.M^I(\vec{k}_1,\vec{k}_2,\lam)
M^I(\vec{k}_3,\vec{k}_4,\lam)\right|_{\lam^2=-\vec{k}^2} \ .
\fq
This factorization at strong coupling is a notable feature in
field theory, given that in general, unitarity cuts in perturbation
theory introduce multi-particle phase space integrals over intermediate
states (momenta, helicity, spin); these integrations generically
do not result in sums of factorized products of
functions.\footnote{We would like to thank
Zvi Bern for numerous discussions on this point.}

\section{Implications for $N=4$ SYM}
\setcounter{equation}{0}

We would like to point out how
the logarithmic dependence in the $\langle \prod_{j=1}^4 {\rm Tr}
F^{\mu\nu} {\tilde F}_{\mu\nu}(k_i) \rangle$ correlator might arise.
First, generically in perturbation theory one encounters $Li_{l+1}(x)$
functions, i.e. polylogarithms,
\bq
Li_k(x) = - \int^x_0 dt ~ {Li_{k-1}(t)\over
t}~,~~~~~~~~Li_0(x)=\frac{x}{(1-x)}~,
\fq
at each $l$-loop order.
For example, in $d$-dimensions at one-loop all $n$-point
functions may be algebraically reduced onto a basis of
integral functions present in $p$-point functions with $p\leq d$
\cite{reduce}.  In four dimensions all of these
integral functions have been computed and contain only rational
functions, logarithms, and dilogarithms.  At higher-loop  certain
classes of Feynman diagrams have been computed; however, one may
understand the logarithmic dependence by examining the cut structure.
For example, consider taking a two-particle cut of a double box
function in the $s$-channel, which separates the double box into a
product of a four-point tree and four-point loop.  The phase space
integral of the (off-shell, i.e.
$k_i^2\neq 0$) cut double-box involves a two-body phase space
integration over a rational function times a function involving
dilogarithms.  Such integrations, in principle, produce $Li_3(x)$
functions.  One may approximately understand the complicated
cut structure of higher loop Feynman diagrams in this manner.

The fact that at large coupling, the imaginary part in a two-particle
channel involves only a product of functions possessing at
most logarithms has the immediate prediction that the result may not
arise at one-loop in field theory.  This is because there are
no one-loop integral functions possessing a $s_{ij}$-channel cut that
can produce squares of logarithms (as they do not cancel in 
preceeding sections).  From the above 
computations it is clear that the 
presence of logarithm squared terms is generic; it appears  
then that AdS four-point correlators of fields dual to bilinear 
(chiral primary and descendent) operators do not reflect 
free-field computations from the CFT-side given this property. 

The interpretation of the strong coupling result within the field
theory context requires that either: the degree $Li_k$
functions conspire to cancel at every order in perturbation theory, or 
that the resummation of the large number of loops add in a
dramatic fashion (after resumming and expanding in the inverse 't Hooft
coupling constant $\lambda=g^2_{YM}N_c$).  An explicit demonstration
of either of these possibilities is difficult to present at the
moment and deserves further study.

\section{Conclusion}

In this work we have computed all of the integrals
necessary for a complete evaluation of the imaginary
parts of the general $N=4$ current correlator at strong coupling
in the 't Hooft limit.  In the Lorentzian signature formulation of the
correspondence, an $i\eps$ prescription is provided that
permits the cutting of the scalar and graviton propagators.
In this work the graviton is analyzed in the non-covariant
$h_{\mu 0}=0$ gauge; however, other gauge choices
lead to similar results.

Explicit expressions are given here for the imaginary
parts in the two-particle channels in the boundary theory momentum
space.  There are two noteworthy aspects of these results.  First,
the imaginary part for the $\langle \prod_{j=1}^4 {\rm Tr}
F^{\mu\nu} \tilde{F}_{\mu\nu} (\vec{k}_j) \rangle$ correlator, as 
well as others, 
at strong 't Hooft coupling factorizes into a product of identical
functions.  Generically (and in any gauge choice)
 any correlator will factorize at strong effective coupling into a
sum of products of functions due to the correspondence with
the classical supergravity formulation.  Second, after explicitly
computing the integrals we see that the final expressions contain
only logarithms and rational functions in the kinematic invariants.
This feature is difficult to see in pure $N=4$ super-Yang-Mills
theory, as generically polylogarithms of any degree are expected.  The
expressions derived in this work indicate large orders of
cancellations of such logarithmic functions. 

The simplicity of the final expressions suggest that an
additional non-trivial structure besides the superconformal
constraints might be appearing in the $N=4$ super
Yang-Mills theory at strong coupling (e.g. \cite{well}).  It would be
interesting to find out if there were any further symmetries of the
classical field equations of gauged supergravity on anti-de Sitter
space; the associated conserved charges would induce those on the
boundary theory and give rise to further Ward identities (possibly
explaining the simple momentum space integral results obtained in this
work). Such symmetries are known to exist in several sets of field
theory equations of motion: in the context of recursively generated
conserved currents of self-dual Yang-Mills field equations,
and Liouville theory which is classically equivalent to a free
field theory.

Further work in field theory requires in the least higher order 
calculations associated with the four-point correlators.  A
calculation of the $\langle \prod_{j=1}^4 {\rm Tr}
F^{\mu\nu} {\tilde F}_{\mu\nu}(k_i) \rangle$ correlator in $N=4$ field
theory would shed light, for example, on how our results do not
agree with free-field $N=4$ super-Yang-Mills theory.  Free-field
expressions agree at the three-point function level for correlators of
chiral primary operators (and their descendants); however,
conformal invariance constrains these functions tightly (up 
to a finite
number of constants).  Our results indicate that generically 
for any four-point AdS boundary correlator the imaginary part 
will contain squares of logarithms (with specific examples 
presented in this work).  Such terms cannot be produced 
in field theory at one-loop and thus the 
four-point functions indeed carry non-trivial information about 
the dynamics of $N=4$ super-Yang-Mills theory,
where the kinematic structure is not constrained by conformal
invariance alone.  This would be evident from a correlation function 
of bilinear operators where the 
free-field result is one-loop.  We should still emphasize that 
within the AdS/CFT correspondence these
functions are relatively simple and factorize in the two-particle
channels at strong coupling.  Similar features are not expected
in $1/N_c$ corrections arising from string-loop effects in the AdS
background and require intermediate phase space integrations.

A complete evaluation of the correlator is also accessible with
the techniques we have generated in this work.  Most of the
necessary integrals have been computed and only the ones arising
from four-point vertices need to be added.
The latter integrals have been computed in position space.  Lastly,
possible connections to the Regge limit of $N=4$ super-Yang-Mills
theory would be worth exploring considering the simple results
\cite{regge}.

\subsection*{Acknowledgements}

We would especially like to thank Zvi Bern for useful and relevant
discussions.  Furthermore, we would like to thank G.\ Bodwin,
F.\ Larsen, H.\ Liu, H.\ Nastase, P.\ van Nieuwenhuizen, M.\ Ro\^cek,
R.\ Siebelink, W.\ Siegel and C.\ Zachos.  G.C. would like to thank the
Aspen Center for Physics for its generous hospitality, during which
time much of this work was carried out.  The work of G.C. is
supported in part by NSF grant No. PHY 9722101 and the U.S.
Department of Energy, Division of High Energy Physics,
Contract W-31-109-ENG-38.

\medskip  

\noindent Note added: That the large $N_c$ limit of four-point
functions of gauge invariant operators in $N=4$ SYM (at finite 
$\lambda$) is indeed not given by
free-field theory has consequently been shown in \cite{fran, eden}.

\end{document}